\documentclass[aps,showpacs,twocolumn,superscriptaddress]{revtex4}

\usepackage{amsmath}
\usepackage{amssymb}
\usepackage{amsfonts}
\usepackage{wasysym}
\usepackage[bookmarks,bookmarksnumbered]{hyperref}

\usepackage[dvips]{graphicx}
\DeclareGraphicsExtensions{.eps}

\newcommand{\be}{\begin{equation}}
\newcommand{\ee}{\end{equation}}
\newcommand{\bea}{\begin{eqnarray}}
\newcommand{\eea}{\end{eqnarray}}
\newcommand{\nn}{\nonumber}

\newcommand{\ket}[1]{|#1\rangle}

\begin{document}

% --- front matter ---------------------------------------------

\bibliographystyle{apsrev}

\title{Quantum dynamics of Bose-Einstein condensates in 
tilted and driven bichromatic optical lattices}

\author{D. Witthaut}
\affiliation{Network Dynamics Group, 
          Max-Planck Institute for Dynamics and Self-Organization, 
          D--37073 G\"ottingen, Germany}
\affiliation{QUANTOP, The Niels Bohr Institute, University of Copenhagen,
 		DK--2100 Copenhagen, Denmark}     
\affiliation{Fachbereich Physik, TU Kaiserslautern,
          D--67663 Kaiserslautern, Germany}			
\author{F. Trimborn}
\affiliation{Institut f\"ur theoretische Physik, Leibniz Universit\"at Hannover, 
          D--30167 Hannover, Germany}    
\affiliation{Fachbereich Physik, TU Kaiserslautern,
          D--67663 Kaiserslautern, Germany}	
\author{V. Kegel}
\affiliation{Fachbereich Physik, TU Kaiserslautern,
          D--67663 Kaiserslautern, Germany}		   
\author{H. J. Korsch}
\affiliation{Fachbereich Physik, TU Kaiserslautern,
          D--67663 Kaiserslautern, Germany}
     
\date{\today }

\begin{abstract}
We study the dynamics of Bose-Einstein condensates in tilted 
and driven optical superlattices. For a bichromatic lattice, each Bloch 
band split up into two minibands such that the dynamics is governed 
by the interplay of Bloch oscillations and transitions between the 
bands. Thus, bichromatic potentials provide an excellent model 
system for the study of nonlinear Landau-Zener tunneling and
allow for a variety of applications in matter wave interferometry
and quantum metrology. In the present paper we investigate the 
coherent dynamics of an interacting Bose-Einstein condensate
as well as its stability. Different mechanisms of instability are 
discussed, which lead to a rapid depletion of the condensate.
\end{abstract}


\maketitle

\section{Introduction}

Despite its apparent simplicity, the dynamics of quantum particles in 
periodic structures is full of surprises. Contrary to our intuition, 
a weak external field inhibits quantum transport in a periodic potential 
in favor of the celebrated Bloch oscillations \cite{Bloc28,Daha98}. 
For stronger fields a directed motion is re-introduced by repeated 
Landau-Zener transitions to higher Bloch bands
\cite{Land32b,Zene32,Majo32,Stue32}.
One of the most interesting physical systems to explore the dynamics in 
periodic potentials are Bose-Einstein condensates  (BEC) in optical lattices, 
allowing an in situ detection of the atoms \cite{Legg01,Mors06}. In recent 
years it became possible to realize periodic potentials with almost arbitrary 
shapes and an astonishing precision. Bichromatic lattices have been 
implemented by superimposing two incoherent optical lattices 
\cite{Gorl01,Foll07}, or by combining optical potentials based on virtual 
two-photon and four-photon processes \cite{Ritt06,Salg07,Salg08,Salg09,Klin10}. 
These superlattices allow to engineer the Bloch band structure of 
the system by tuning few experimental parameters.

The dynamics of a BEC in a tilted optical lattice, and especially 
the Landau-Zener tunneling between  Boch bands, is strongly 
modified by the inter-atomic interactions. While the fundamental 
problem of Landau-Zener tunneling between two levels was solved 
as early as 1932 independently by Landau, Zener, Majorana and St\"uckelberg \cite{Land32b,Zene32,Majo32,Stue32}, a generalization of these 
results to interacting many-particle systems remains an open question 
up to today. The possibility to investigate Landau-Zener tunneling of
a BEC in situ in a well controllable laboratory experiment has thus 
attracted much interest in recent years.
A distinguished result of these studies was that strong interactions 
in a BEC lead to a breakdown of adiabaticity and instability even in 
the limit of very slow parameter variations.
This phenomenon was first predicted theoretically within a 
mean-field approximation \cite{Wu00,Zoba00,Liu02,05level3} 
and demonstrated for a BEC in an accelerated optical lattice 
shortly afterwards \cite{Ande98,Lasi03,Fall04,Sias07,Zene09}.
Later, it was shown theoretically that the breakdown of adiabaticity 
results from the occurence of diabatic level crossings in the 
many-body spectrum \cite{06zener_bec,Wu06,10zener_phase}.

In a bichromatic optical lattice, it is possible to tune the energy 
gap between the minibands via the relative phase of the two lattices
and thus to control Landau-Zener tunneling as shown in \cite{Salg07}.
Furthermore, the band stucture can be engineered such that
the BEC is confined to the two lowest minibands and cannot be 
lost by repeated tunneling to higher excited 
bands, such that one can even observe repeated Landau-Zener 
tunneling events \cite{06bloch_zener,06bloch_manip,Klin10}.
In addition, Landau-Zener tunneling can be used as a coherent
beam splitter for atomic matter waves, which enables a variety 
of possible applications in matter wave interferometry and
quantum metrology \cite{Klin10}. 

In the present paper we will provide a thorough theoretical analysis 
of nonlinear Landau-Zener tunneling in bichromatic optical lattices
with a focus on the onset of dynamical instability and depletion
of the condensate.
Furthermore, we study Bloch-Zener oscillations, the coherent 
superposition of Bloch oscillations and Landau-Zener tunneling, 
and the dynamics in driven lattices, where transitions between 
the minibands are caused by the periodic driving instead of a
static field. In the latter case, the dynamics of
a BEC is governed by the interference of two matter waves 
with the same quasi momentum and opposite group velocity.
This gives rise to an oscillatory motion which can be viewed as
a simple quantum simulation of the {\it Zitterbewegung} of a
relativistic Dirac spinor.

We focus on a bichromatic optical lattice with
an alternating depth of the lattice wells. The many-body dynamics in such
an optical lattice is described by the Bose-Hubbard type hamiltonian 
\cite{Jaks98}
\bea 
   H &=& - J \sum_{n=1}^{M-1} \left( \hat a_{n+1}^\dagger \hat a_n 
         + \hat a_n^\dagger \hat a_{n+1} \right)
    + \frac{U}{2} \sum_{n=1}^{M}  \hat a_n^{\dagger 2} \hat a_n^{2} \nn \\
   &&  \qquad + \sum_{n=1}^{M}  \left( \frac{\delta}{2} (-1)^n + Fn \right)
       \hat a_n^\dagger \hat a_n,
    \label{eqn-ham-bh-2band}
\eea
where $\hat a_n$ and $\hat a_n^\dagger$ are the bosonic
annihilation and creation operators, respectively. The parameter
$J$ denotes the tunneling rate between the wells, $U$ is 
the interaction strength and $F$ the strength of the static external 
field which accelerates the atoms. The parameter $\delta\ge0$ is the
difference of the on-site energies between adjacent wells. It
is directly proportional to the intensity of the double-periodic
optical lattice. In the following we set $J = 1$ in all simulations,
i.e. we measure all energies in units of the tunneling matrix
element $J$.

One of the main objectives of the present paper is a careful
analysis of how the interactions affect the coherent dynamics
of a BEC in a bichromatic lattice and how they possibly lead
to instabilities of the condensate. Throughout the paper we thus
assume that the system is initially prepared as a pure BEC
with $N$ particles:
\be
  \ket{\Psi(0)} = \frac{1}{\sqrt{N!}} 
     \left( \sum \nolimits_n \psi_n \hat a_n^\dagger \right)^N \ket{0}.
  \label{eqn-2band-mp-initial}
\ee
As long as the quantum state remains close to a pure BEC, the 
dynamics is well described within a mean-field approximation. The 
dynamics of the condensate wave function is then given by the 
discrete nonlinear  Schr\"odinger equation (DNLSE) \cite{Mors06}
\be  
   i \dot \psi_n = -J(\psi_{n+1}+\psi_{n-1}) + 
     \left[  \frac{\scriptstyle (-1)^n \delta}{\scriptstyle 2}  
       + Fn  + UN |\psi_n|^2 \right] \psi_n.
     \label{eqn-dnlse}    
\ee
However, dynamical instabilities lead to a depletion of the condensate
such that the mean-field approximation is no longer applicable.
In order to simulate  the dynamics beyond mean-field we use 
the Bogoliubov backreaction (BBR) method which also gives
a quantitative prediction for the depletion of the condensate
\cite{Vard01,Angl01,Tikh07}.

\section{Bloch states and bands}

Bloch states are the simultaneous eigenstates of the field-free 
Hamiltonian and a translation over \emph{two} lattice sites. 
The dynamics of a BEC can be understood to a large extent 
from the properties of linear and non-linear Bloch states,
including the depletion of the condensate. In the following 
we thus give a detailed analysis of the Bloch states and their 
stability.

\subsection{Single-particle Bloch states}

In the simplest case of a single atom, where interactions are obviously 
irrelevant, the Bloch bands are easily calculated as \cite{06bloch_zener}
\be
   E_{\alpha}(\kappa) = \frac{(-1)^{\alpha+1}}{2}
      \sqrt{\delta^2 + 4J^2 \cos^2(\kappa)},
\ee
where $\alpha = 0,1$ labels the two minibands and 
$\kappa \in [-\pi/2, +\pi/2]$ denotes the quasimomentum. The 
band gap between the two minibands is directly given by the 
parameter $\delta$. The corresponding Bloch states are given by 
$| \chi_{\alpha,\kappa} \rangle = \hat b_{\alpha,\kappa}^\dagger |0\rangle$,
where
\bea
    \hat b_{0,\kappa} &=& \frac{1}{\sqrt{N_\kappa}} \sum_n
             u_\kappa e^{i 2n \kappa } \hat a_{2n} +
             v_\kappa e^{i (2n+1) \kappa} \hat a_{2n+1} \nn \\
    \hat b_{1,\kappa} &=& \frac{1}{\sqrt{N_\kappa}} \sum_n
             v_\kappa e^{i 2n \kappa } \hat a_{2n} -
             u_\kappa e^{i (2n+1) \kappa} \hat a_{2n+1}    
    \label{eqn-band-lin}
\eea
and
\bea
  u_\kappa &=& 4J \cos(\kappa) \nn \\ 
  v_\kappa &=& \delta + \sqrt{\delta^2 + 4J^2 + \cos^2(\kappa)}, \nn
\eea
$N_\kappa = \pi(u_\kappa^2 + v_\kappa^2)$ being a normalization
constant.

\subsection{Nonlinear Bloch bands}

\begin{figure}[t]
\centering
\includegraphics[width=8cm,  angle=0]{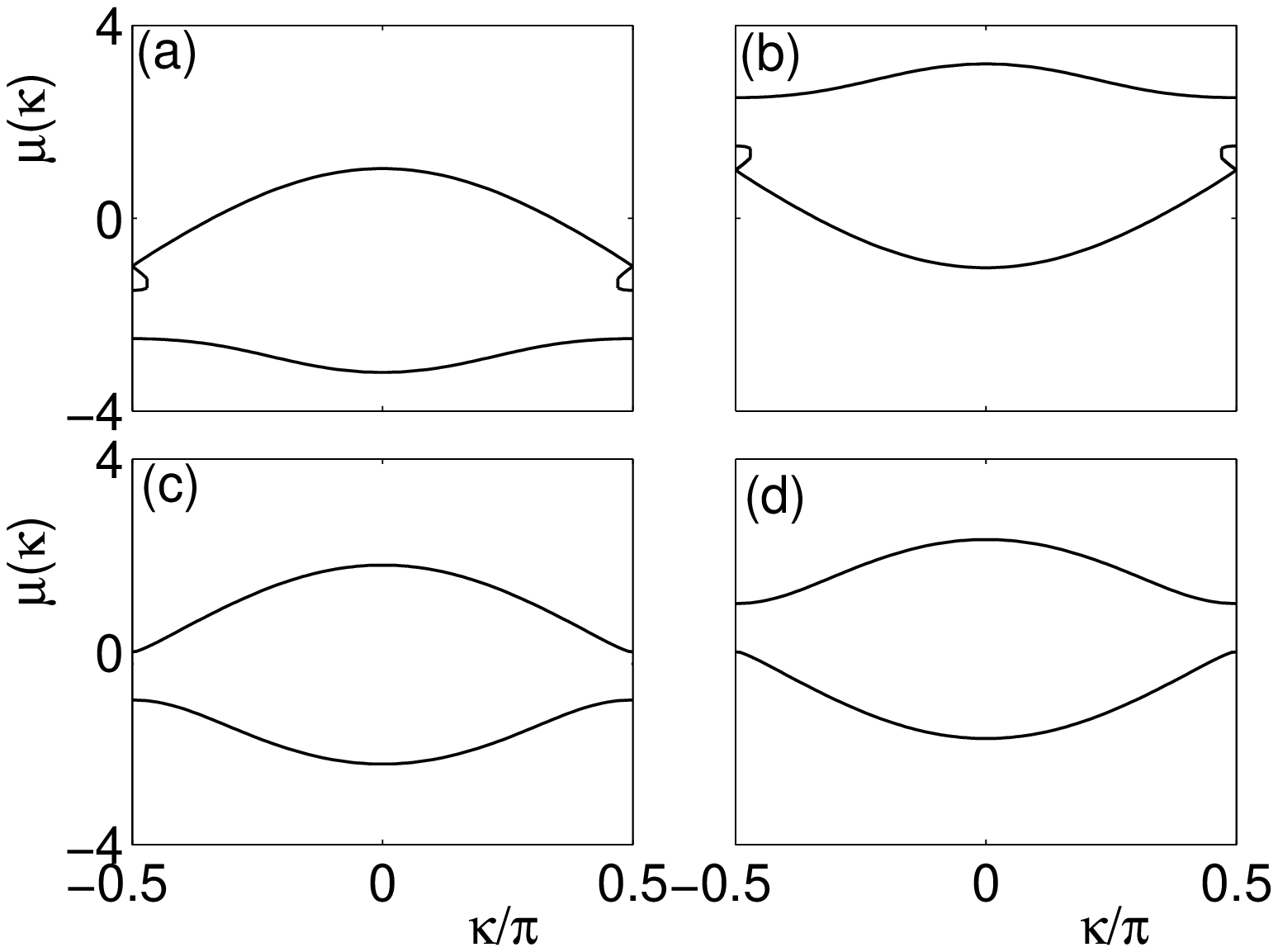}
\includegraphics[width=5cm,  angle=0]{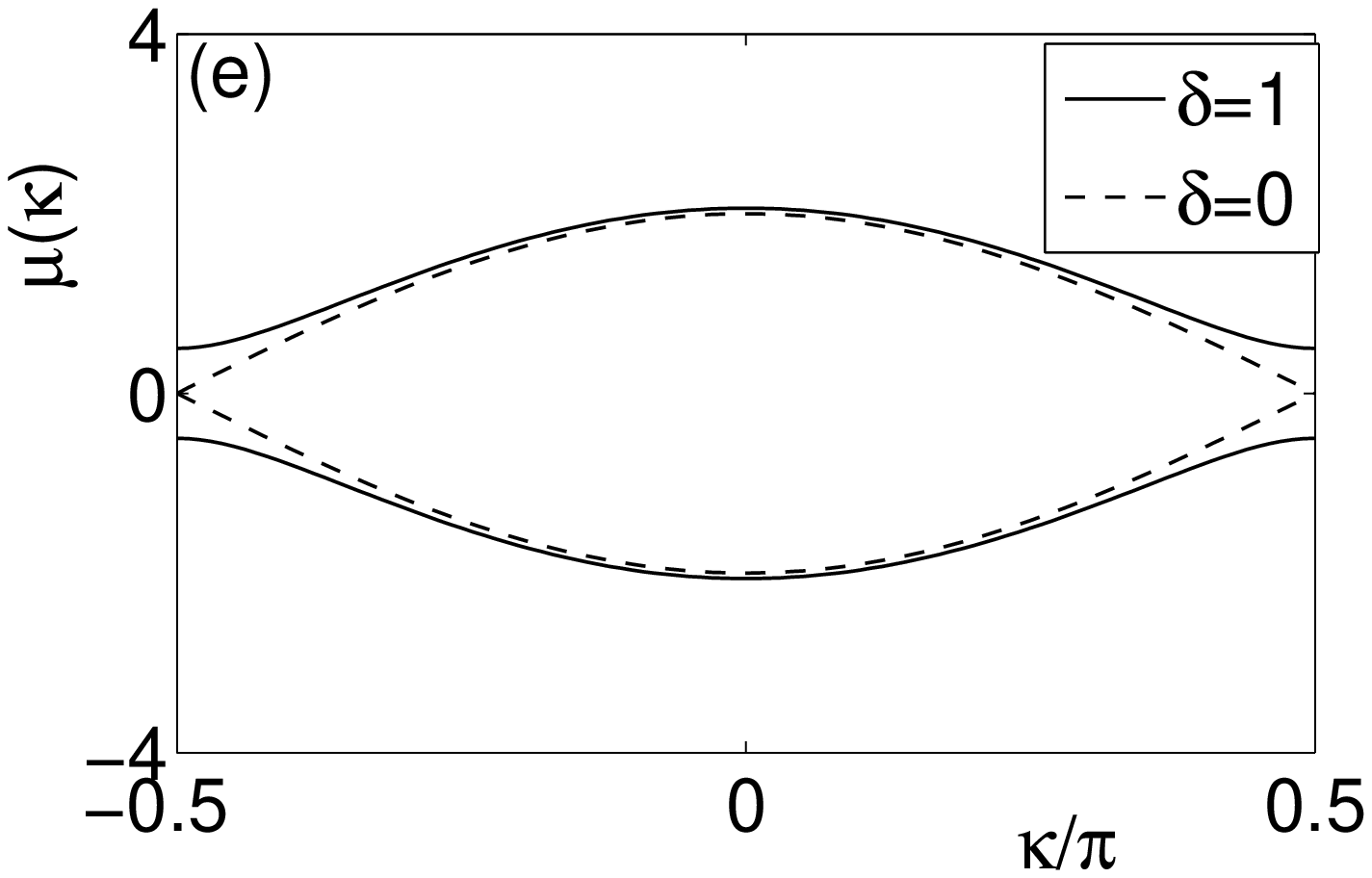}
\caption{\label{fig-blochbands}
Nonlinear Bloch bands in a bichromatic optical lattice with $\delta = 1$ 
for $g = -2$ (a), $g = +2$ (b), $g=-0.5$ (c) and $g=+0.5$ (d). 
Linear Bloch bands ($g=0$) are shown in (e) for $\delta = 1$
(solid line) and $\delta = 0$ (dashed line) for comparison.
 }
\end{figure}

Also in the mean-field approximation one can analytically calculate the 
'nonlinear' Bloch states, which are defined as stationary states of the
DNLSE 
\be
   \mu \phi_n = -J (\phi_{n+1} + \phi_{n-1}) + UN |\phi_n|^2 \phi
      + (-1)^{n}  \frac{\delta}{2} \phi_n,
\ee
with the translation symmetry $\phi_{n+2} = e^{2i \kappa} \phi_n$.
Making the ansatz
\be
  \phi_n \sim \left\{ \begin{array}{l l}
     u_\kappa e^{i \kappa n} \; & n \, \mbox{even} \\
     v_\kappa e^{i \kappa n} \; & n \, \mbox{odd}, \\
   \end{array}  \right.
   \label{eqn-BSnonlin}
\ee
one finds that the coefficients $u_\kappa,v_\kappa$ are determined
by the two-mode DNLSE 
\be
  \left( \begin{array} {c c}
    \delta/2 + g |u_\kappa|^2 & -2J \cos(\kappa) \\
    -2J \cos(\kappa) & -\delta/2 + g |v_\kappa|^2
   \end{array} \right) 
   \left(\begin{array}{c} u_\kappa \\ v_\kappa \end{array} \right)
   = \mu  \left(\begin{array}{c} u_\kappa \\ v_\kappa \end{array} \right).
   \label{eqn-gpe-2mode}
\ee
Using the normalization $|u_\kappa|^2 + |v_\kappa|^2 = 1$, 
the effective coupling constant is given by $g = 2U\rho$, $\rho$
being the average particle density. Examples of nonlinear 
Bloch bands are shown in  Fig.~\ref{fig-blochbands}.
One observes that the bands become strongly asymmetric --
for a repulsive nonlinearity $g>0$  the curvature of the ground 
band increases while the curvature of the excited band decrease 
and vice versa for an attractive nonlinearity $g <0$.
For strong nonlinearities novel stationary states appear
at the band edge $\kappa = \pm \pi/2$, forming the so-called 
looped Bloch bands \cite{Wu03,09dcomb}. A quantitative analysis 
shows that DNLSE (\ref{eqn-gpe-2mode}) admits four solutions 
if \cite{Liu02}
\be
   g^{2/3}  > \delta^{2/3} + |2 J \cos(\kappa)|^{2/3}.
\ee
Thus the critical nonlinearity for the existence of looped
levels is directly linked to the band gap $\delta$. The 
deformation of the Bloch bands has significant consequences 
for the dynamics which will be discussed in detail in 
Sec.~\ref{sec-zener-tunneling}. An adiabatic dynamics
is hindered by a sharpening of the levels and becomes 
completely impossible as soon as the loops form.

\subsection{Stability analysis}
\label{sec-stability}

\begin{figure}[t]
\centering
\includegraphics[width=8cm,  angle=0]{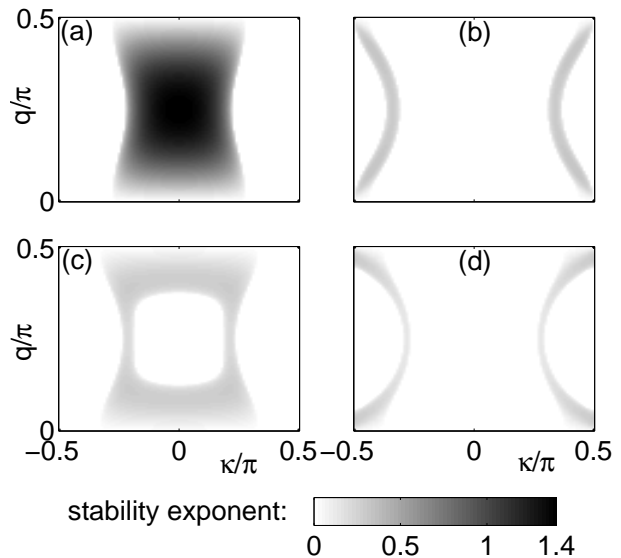}
\caption{\label{fig-stabmap}
Stability map for a nonlinear Bloch state with quasimomentum
$\kappa$ in the ground band for $\delta = 1$ and $g = -2$ (a), 
$g = +2$ (b), $g=-0.5$ (c) and $g=+0.5$ (d). The gray scale
map shows the stability exponent, i.e. the growth rate of a
perturbation with quasimomentum $q$. Dynamical instability
leads to a depletion of the condensate if the stability exponent
is non-zero for at least one value of $q$.
}
\end{figure}

The nonlinear Bloch states calculated in the preceding section
are stationary states of the DNLSE (\ref{eqn-dnlse}). However,
they can become dynamically unstable due to the nonlinear
interaction term, which also indicates a rapid depletion of
the BEC. In order to determine the stability properties 
of a Bloch state (\ref{eqn-BSnonlin}), we add a small perturbation 
with quasimomentum $q$:
\bea
      \label{eqn-stab-ansatz}  
   && \psi_n(t) = e^{- i \mu t} \phi_n \\
   && \qquad \qquad + \left\{ \begin{array}{l l}
     e^{i(\kappa n - \mu t)} (\xi_{0,\kappa} e^{iqn} + \zeta_{0,\kappa}^* e^{-iqn} ) &
          \; n \, \mbox{even} \\
     e^{i(\kappa n - \mu t)} (\xi_{1,\kappa} e^{iqn} + \zeta_{1,\kappa}^* e^{-iqn} ) &
          \; n \, \mbox{odd}  \\
     \end{array} \right. \nn  
\eea
and analyze the consequences for the energy and the dynamics
of the condensate. If every perturbation increases the 
total energy given by the Gross-Pitaevskii energy functional 
\bea
   E &=&  -J \sum_n \psi_{n+1}^* \psi_n + \psi_n^* \psi_{n+1}
            +  \frac{UN}{2} \sum_n |\psi_n|^4  \nn \\
     && \qquad    + \frac{\delta}{2} \sum_n (-1)^2 |\psi_n|^2,
   \label{eqn-gpe-energy}
\eea
then the Bloch state represents a local energy minimum and thus 
a stable superflow. Otherwise a perturbation may lower the energy 
and the Bloch state suffers a Landau instability.
In the present paper we are more concerned with the dynamical
instability of a Bloch state, which occurs if a perturbation grows
exponentially, as this indicates a rapid depletion of the condensate
mode \cite{Cast98,07phase,09phase_appl,Kolo09}. 
Note that dynamical instability always indicates
energetical instability but not vice versa \cite{Wu03}.

In order to determine the energetical stability we substitute the
ansatz (\ref{eqn-stab-ansatz}) into the Gross-Pitaevskii energy
functional (\ref{eqn-gpe-energy}) and expand it up to second order
in the perturbation. The variation of the energy is then given by
\be
  \delta E = \int dq \;
     \Xi_\kappa^\dagger      L_{\rm en}(\kappa,q) \Xi_\kappa
\ee
with the matrix
\be
   L_{\rm en}(\kappa,q) = \left( \begin{array}{c c}
   H(\kappa+q) & gV \\
   g V^* & H(\kappa-q) 
   \end{array} \right). 
\ee
Here we have introduced the abbreviations
\bea
   H(k)  \!\! &=& \!\! \left( \begin{array}{c c}
   \frac{\delta}{2}+2g |u|^2 - \mu  & -2J \cos(k)  \\
   -2J \cos(k)  & -\frac{\delta}{2}+2g |v|^2 - \mu \\
   \end{array} \right), \nn \\
   V\!\!  &=& \!\!  \left( \begin{array}{c c} 
   u^2 & 0 \\ 0 & v^2 
   \end{array} \right) \qquad \mbox{and} \nn \\
   \Xi_\kappa \!\!  &=& \!\! (\xi_{0,\kappa},\xi_{1,\kappa},
         \zeta_{0,\kappa},\zeta_{1,\kappa})^T.
\eea
The Bloch state represents a stable energy minimum if $\delta E$
is positive for any perturbation, i.e. if the matrix $L_{\rm en}(\kappa,q)$ 
is positive definite for every $q$.

The dynamical stability properties are found by substituting the
ansatz (\ref{eqn-stab-ansatz}) into the DNLSE  (\ref{eqn-dnlse}). 
In first order, the perturbation evolves according to the 
Bogoliubov-de Gennes equation
\be
   i \frac{d}{dt} \Xi_\kappa    = L_{\rm BdG}(\kappa,q) \Xi_\kappa
\ee
with
\bea
   L_{\rm BdG}(\kappa,q) &=& \sigma_z  L_{\rm en}(\kappa,q) \nn \\
   &=& \left( \begin{array}{c c}
   H(\kappa+q) & gV \\
   -gV^* & -H(\kappa-q) 
   \end{array} \right).
\eea
A dynamical instability occurs if a perturbation grows exponentially,
i.e. if there is any $q$ for which the eigenvalues of the matrix
$L_{\rm BdG}(\kappa,q)$ are not purely real.

The dynamical stability for the Bloch states in the ground band is 
depicted in Fig.~\ref{fig-stabmap} for the same parameters as in 
Fig.~\ref{fig-blochbands}. A grey scale map shows the stability
exponent, i.e. the maximum imaginary
part of the eigenvalues of the Bogoliubov-de Gennes matrix
$L_{\rm BdG}(\kappa,q)$ in dependence of $\kappa$ and $q$.
This imaginary part  indicates the growth rate of a perturbation 
with wavenumber $q$ and thus also the depletion rate of the 
condensate. A Bloch state with quasimomentum $\kappa$ is 
dynamically stable only if the growth rate is zero for all values 
of $q$. 
One observes that two different kinds of dynamical instability exist
in a bichromatic lattice. In the attractive case $g <0$, the Bloch 
states in the center of the Brillouin zone become strongly unstable
already for a quite modest nonlinearity. Thus we face the surprising
conclusion that an attractive interaction on the one hand flattens the 
ground band and thus faciliates an adiabatic evolution, but on the 
other hand leads to instability.  
On the contrary, a strong repulsive nonlinearity is required to 
introduce a weak dynamic instability at the edge of the Brillouin 
zone, which is associated with the occurence of looped
Bloch bands. Thus one can infer that a significant 
depletion takes place only for a much stronger interaction than 
in the attractive case and that it sets in at the edge of the Brillouin zone
around $\kappa = \pi/2$.

\section{Dynamics}

\begin{figure}[t]
\centering
\includegraphics[width=8.5cm,  angle=0]{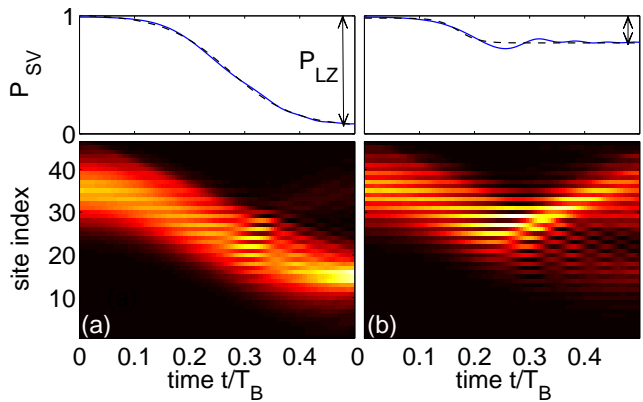}
\caption{\label{fig-mplz-2band-ex}  
(Color online)
Many-particle Landau-Zener tunneling between two minibands 
with band gap $\delta = 0.2$ (a) and $\delta = 1$ (b). The dynamics 
has been calculated with the BBR method for the parameters 
$J=1$, $F = 0.2$,  $UN = 2$ and $N=100$ particles. The upper 
panels show the survival probability (\ref{eqn-psv}) in the upper 
half of the lattice (solid line).}
\end{figure}

The dynamics of a BEC in a bichromatic optical lattice is governed
by the interplay of the intraband dynamics and the transitions between 
the two minibands. 
A static force can be introduced by gravity \cite{Ande98}, magnetic 
field gradients \cite{Gust08}, or by an acceleration of the complete 
lattice (see \cite{Mors06} and references therein).
This force accelerated the atoms until they are finally Bragg reflected, 
leading to the celebrated Bloch oscillations \cite{Bloc28,Daha98,Salg08}.
In addition, this force induces Landau-Zener tunneling between
the minibands \cite{Salg07}, which will be investigated in detail in the following.
In a bichromatic lattice one can control the tunneling rate to a
large extent by tuning the lattice parameters \cite{Ritt06}.
On longer timescale repeated Landau-Zener tunneling takes place,
which leads to a complex dynamics due to the interference effects
of atoms in the two minibands \cite{06bloch_zener,06bloch_manip,Klin10}. 
Finally, we will extend our analysis
to the case where transitions are not driven by an external field but
by a periodic driving.

In order to simulate the dynamics beyond the mean-field approximation 
and to describe the depletion of the condensate we use the 
Bogoliubov backreaction (BBR) method introduced in 
\cite{Vard01,Angl01,Tikh07}. 
In this approach one takes into account two- and four point functions 
and truncates all higher order correlation functions to obtain a 
closed set of evolution equations. The BBR 
method has proven its worth to predict the features of the many-body
quantum state, especially the depletion of the condensate mode, 
avoiding the common problems of the Hartree-Fock-Bogoliubov 
approximation \cite{Vard01,Angl01,Tikh07}. In particular the nature of the 
many-body quantum state is indicated by the reduced single-particle density 
matrix (SPDM) \cite{Vard01,Angl01,Tikh07,Legg01} 
\be
  \sigma_{j,m} = \frac{1}{N} \langle \hat a_j^\dagger \hat a_m \rangle.
\ee  
The leading eigenvalue of this matrix, $\lambda_0$, gives the fraction
of atoms in the condensate mode. Consequently, the non-condensed
fraction is given by $1-\lambda_0$.

\subsection{Nonlinear Zener tunneling}
\label{sec-zener-tunneling}

To begin with, we explore the basic features of the dynamics in a 
bichromatic lattice subject to a static external field, Bloch oscillations and 
Landau-Zener tunneling, for the weakly interacting case. 
We assume that the initial state is a pure BEC (\ref{eqn-2band-mp-initial})
in the ground Bloch band with $\kappa = 0$ weighted by a 
Gaussian envelope $\psi_n(0) \sim \phi_n \exp (-(n-n_0)^2/4\sigma^2)$
with a width of $\sigma = 5$ sites centered 
around the site  $n_0 = 35$.  In the forthcoming examples we choose 
the total particle number to be $N=100$ located in a lattice with 
$M=46$ sites and $F=0.2$, unless otherwise stated.
A weak nonlinearity induces a reversible dephasing, which damps
Bloch oscillations \cite{05bloch_bec,Gust08,Kolo09}, and, of course, alters 
the Landau-Zener tunneling rate between the two minibands.

Figure \ref{fig-mplz-2band-ex} shows two examples of the many-body 
dynamics starting from the initial state (\ref{eqn-2band-mp-initial}) 
for a weak repulsive interaction, $UN = 2$, and two 
different values of the band gap $\delta$. The figures show the 
evolution of the density $\langle \hat n_j(t)\rangle, \; j = 1,\ldots,M$
in false color.
Here and in the following, we take the Bloch time $T_B=2\pi/F$ 
of the single-periodic lattice ($\delta=0$) as the reference time scale.
One observes that the BEC is first accelerated by the external 
field $F$ until it reaches the edge of the Brillouin zone at $T=T_B/4$. If the 
band gap $\delta$ is large (Fig.~\ref{fig-mplz-2band-ex} (b)), the 
BEC matter wave stays in the ground miniband and performs 
Bloch oscillations with a period of $T_B/2$.
In contrast, if the gap is small, the matter wave tunnels to the excited 
miniband and performs Bloch oscillations with the full 
period $T_B$ (Fig.~\ref{fig-mplz-2band-ex} (a)). At time $t = T_B/2$ 
it is located at the turning point of the Bloch oscillations. For 
intermediate values of the band gap, only a fraction of the 
condensate tunnels to the excited miniband and the wavepacket 
splits.

\begin{figure}[t]
\centering
\includegraphics[width=8cm,  angle=0]{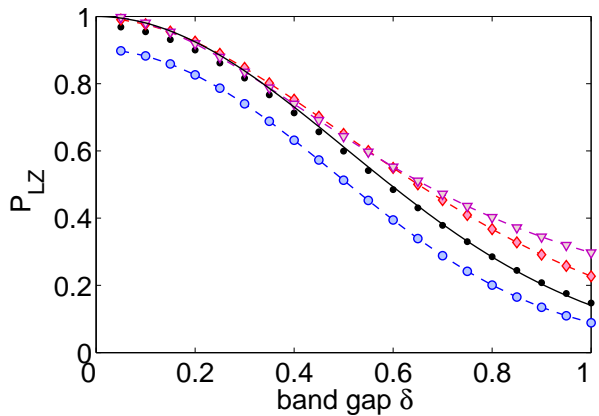}
\caption{\label{fig-mplz-2band-pvsd}  
(Color online)
Landau-Zener tunneling probability $P_{\rm LZ}$ between two 
minibands as a function of the band gap $\delta$ for 
$UN=-2$ ($\circ$), $UN = 0$ ($\cdot$), $UN = +2$ ($\diamond$) 
and $UN = +4$ ($\triangledown$) and $N=100$ particles.
The solid line shows the analytic approximation (\ref{eqn-plz-2band-lin})
for the linear case. Dashed lines are plotted to
guide the eye.
 }
\end{figure}

For a further quantitative analysis of the Landau-Zener tunneling rate 
we estimate the survival probability in the ground miniband
by the number of atoms remaining in the upper half of the lattice:
\be
 P_{\rm SV}(t) = \frac{1}{N} \sum_{j=24}^{46} \langle \hat n_j(t)\rangle.
 \label{eqn-psv}
\ee
The Landau-Zener tunneling probability to the excited band is 
then given by  $P_{\rm LZ} := 1-P_{\rm  SV}(T_B/2)$.
The time dependence of the survival probability (\ref{eqn-psv})
is shown in the upper panels of Fig.~\ref{fig-mplz-2band-ex}  
together with $P_{\rm LZ}$.

\begin{figure}[t]
\centering
\includegraphics[width=8cm,  angle=0]{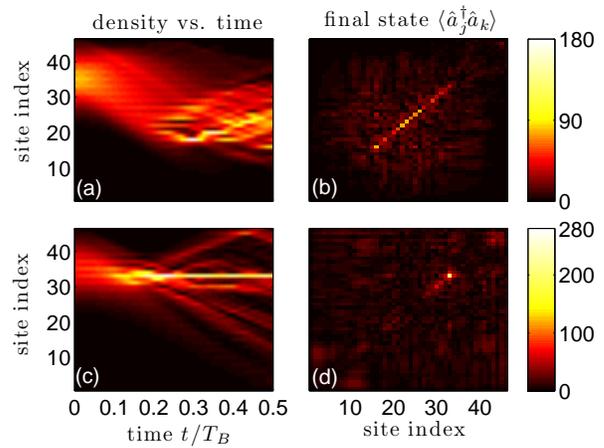}
\caption{\label{fig-mp-instability}  
(Color online)
Unstable dynamics in a tilted bichromatic optical lattice for strong
interactions $UN = +20$ (a,b) and $UN = -10$ (c,d).
The left panels (a,c) show the evolution of the density and the
right-panels (b,d) show the magnitude of the scaled SPDM
$N |\sigma_{j,k}| = |\langle \hat a_j ^\dagger \hat a_k \rangle|$ 
at $t = T_B/2$ in a colormap plot.
The remaining parameters are  $\delta = 0.2$ and $N=1000$.
}
\end{figure}

Figure \ref{fig-mplz-2band-pvsd} shows  the Landau-Zener tunneling 
probability $P_{\rm LZ}$ in dependence of the band gap $\delta$ for 
different values of the interaction strength $UN$. 
In the linear case, $UN=0$, one can approximate the avoided crossing of 
the two minibands at the edge of the Brillouin zone by an effective two-level 
model which yields the following approximation for the Landau-Zener 
probability \cite{06bloch_zener}:
\begin{eqnarray}
  P_{\rm LZ}^{(0)} \approx \exp\left(- \frac{\pi \delta^2}{8J F} \right).
  \label{eqn-plz-2band-lin}
\end{eqnarray}
This approximation shows an excellent agreement with the numerical 
results shown in Fig.~\ref{fig-mplz-2band-pvsd}. In the weakly nonlinear 
case one observes an increase of the Landau-Zener tunneling rate 
$P_{\rm LZ}$ for a repulsive nonlinearity  $U>0$ and a decrease for 
an attractive nonlinearity $U<0$, which has also been demonstrated 
experimentally \cite{Lasi03}. This effect can be understood from the 
structure of the nonlinear Bloch states introduced above. With increasing 
interaction strength, the nonlinear Bloch bands  $\mu(\kappa)$ become 
strongly asymmetric as shown in Fig.~\ref{fig-blochbands}. For $UN<0$, 
the ground band is flattened so that adiabaticity is faciliated and 
$P_{\rm LZ}$ decreases, while the excited band is sharpened.
The inverse effect is found for  $UN > 0$ such that $P_{\rm LZ}$ increases. 
If the nonlinearity $UN$ exceeds a critical value, a Bloch state at the 
edge of a band bifurcates to a looped structure, which prevents an
adiabatic evolution even for very small values of $F$.

\subsection{Depletion of the condensate}

Strong inter-atomic interactions alter the dynamics of the
BEC completely. Examples are shown in Fig.~\ref{fig-mp-instability}    
for a repulsive (a,b) and an attractive (c,d) interaction, respectively.
One observes that the familiar Bloch oscillation pattern is significantly 
disturbed, especially in the case of attractive interactions.
In the repulsive case, the atoms are distributed over several
lattice sites, but the phase coherence between theses sites
is lost almost completely. This is indicated by a strong 
suppression of the non-diagonal parts of the SPDM as 
shown in Fig.~\ref{fig-mp-instability}  (b).
A strong attractive nonlinearity leads to a collapse of the 
condensate. Figure \ref{fig-mp-instability} (c) shows that the 
atoms are strongly focussed to a single lattice site at 
$t \approx 0.15 \, T_B$. Afterwards, a fraction of the atoms 
'explodes' from the focus and the condensate mode is 
rapidly depleted. 

\begin{figure}[t]
\centering
\includegraphics[width=8cm,  angle=0]{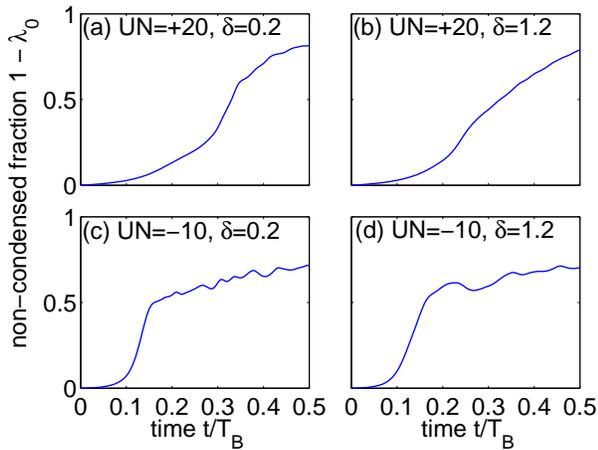}
\caption{\label{fig-mplz-p0vsg}  
(Color online)
Depletion of the condensate during nonlinear Landau-Zener tunneling.
Temporal growth of the non-condensed fraction $1-\lambda_0$ for 
a strong repulsive interaction $UN = + 20$ (a,b) and an attractive 
interaction $UN = -10$ (c,d). The remaining parameters are
$\delta = 0.2$ (a,c) and $\delta = 1.2$ (b,d), respectively,
and $N=1000$.
}
\end{figure}

To further analyze the different mechanisms of instability 
due to repulsive and attractive interactions we calculate how
the condensate is depleted. Figure \ref{fig-mplz-p0vsg} shows
the time evolution of the non-condensed fraction $1-\lambda_0$ 
for a strong repulsive ($UN=+20$) and a strong attractive nonlinearity 
($UN=-10$), respectively. In the attractive case, instability sets 
in much earlier and takes place on a very short time scale.
This difference can be explained by the results of a linear stability
analysis as 
discussed in Sec.~\ref{sec-stability}. In the repulsive case the onset of 
dynamical instability can be associated with the emergence of looped Bloch 
bands. The condensate becomes dynamically unstable at the edge 
of the Brillouin zone where the loops emerge. In contrast, already 
modest attractive interactions lead to a dynamic instability at the 
center of the Brillouin zone  (cf.~Fig~\ref{fig-stabmap}) such that 
the depletion of the condensate sets in immediately.

\begin{figure}[t]
\centering
\includegraphics[width=8cm,  angle=0]{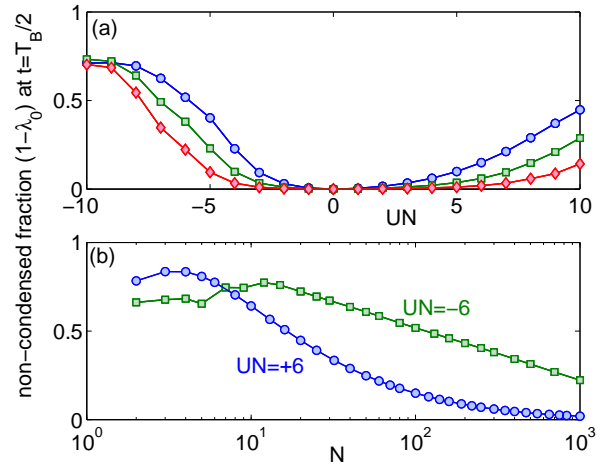}
\caption{\label{fig-mplz-depfinal}  
(Color online)
The non-condensed fraction $1-\lambda_0$ at time $t = T_B/2$ as a 
function of (a) the interaction strength $UN$ for $N=100$ ($\circ$), 
$N=300$, ($\square$) and $N = 1000$ ($\diamond$) and
(b) as a function of the particle number $N$ for a fixed interaction 
strength $UN = +6$ ($\circ$) and $UN = -6$ ($\square$).
Solid lines are drawn to guide the eye.}
\end{figure}

A quantitative analysis of the depletion of the condensate is provided
in Fig.~\ref{fig-mplz-depfinal}, where we have plotted the non-condensed 
fraction at $t = T_B/2$ as a function of the interaction strength
$UN$  in (a) and the particle number $N$ for a fixed value of the
interaction strength $UN = \pm 6$ in (b). 
Figure \ref{fig-mplz-depfinal} (a) clearly shows the qualitative difference
between an attractive and a repulsive interaction.  In the first case,
one observes a rapid increase of the non-condensed fraction when 
the interaction strength exceeds the critical value for the onset of a 
dynamical instability. For a repulsive interaction, however, the dynamics
is rather stable so that the non-condensed fraction remains small
for all values of $|UN| < 10$ shown in the figure.
The non-condensed fraction decreases with the particle number
and tends to zero in the mean-field limit $N \rightarrow \infty$.
However, the speed of convergence depends crucially on the
stability of the dynamics as shown in  Fig.~\ref{fig-mplz-depfinal} (b). 
In the repulsive case, $UN = +6$, the dynamics is stable and thus
convergence is fast. The non-condensed fraction decreases rapidly
with increasing particle number such that the mean-field description 
by the DNLSE is valid already for quite small values of the particle number. 
On the contrary, the convergence is logarithmically slow for $UN = -6$ 
due to the dynamical instability. 
A different approach to a classical instability causing depletion is
provided by generalized mean-field descriptions  
\cite{Cast98,07phase,09phase_appl}.

\subsection{Bloch-Zener oscillations}

\begin{figure}[t]
\centering
\includegraphics[width=8.5cm,  angle=0]{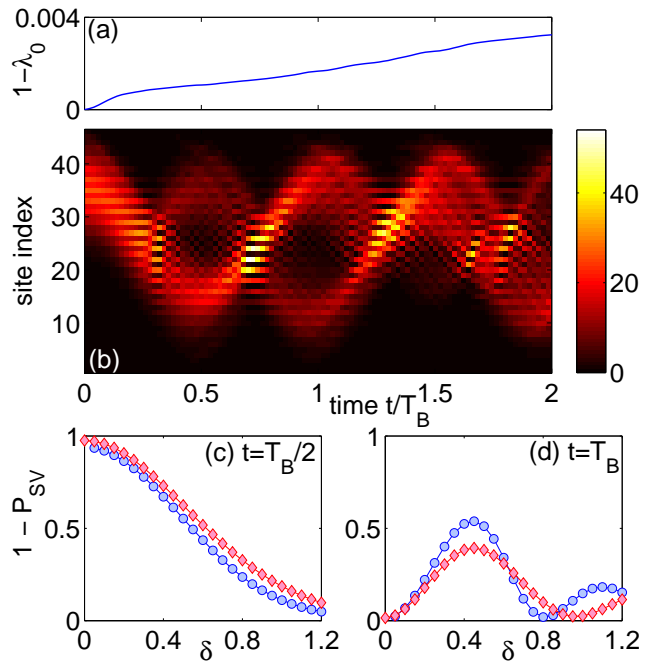}
\caption{\label{fig-bzo1}  
(Color online)
Bloch-Zener oscillations of a BEC with $N=300$ atoms in a 
tilted bichromatic lattice.  
(a) Non-condensed fraction $1-\lambda_0$  for $\delta = 0.5$ and $UN = 1$.
(b) Atomic density in real space for $\delta = 0.5$ and $UN = 1$.
(c,d) Survival probability (\ref{eqn-psv}) in the upper half of the lattice
after $t = T_B/2$ and $t = T_B$, respectively, for $UN = -1$
($\circ$) and $UN = +1$ ($\diamond$).
}
\end{figure}

On a longer timescale, the dynamics of a BEC in a tilted optical lattice
is governed by the interference of Bloch oscillations and Zener tunneling
between the Bloch bands \cite{06bloch_zener,06bloch_manip}.
Figure \ref{fig-bzo1} (b) shows an example of the dynamics of 
the atomic density for $\delta = 0.5$. The condensate wave packet 
is coherently split by Landau-Zener tunneling between the two 
minibands at $t = T_B/4$ and recombined again at $t = 3T_B/4$,
thus realizing an effective matter wave Mach-Zehnder interferometer.
The splitting ratio of this interferometer, which is given by the 
Landau-Zener tunneling rate (\ref{eqn-plz-2band-lin}), is easily 
tunable by changing the band gap $\delta$.

For very small and for very large values of  $\delta$, 
the condensate occupies only one miniband -- it remains in the 
ground band for large $\delta$ and tunnels completely to the other
miniband for small $\delta$ as shown in Fig.~\ref{fig-bzo1} (c).
In both cases the condensate shows simple Bloch oscillations and
returns back to its initial state at $t=T_B$.
For intermediate values of $\delta$, the condensate is split into two 
parts at $t = T_B/4$. The further dynamics and especially the 
occupation of the two minibands is governed by the interference 
of the two possible paths. For the given parameters, about one 
half of the population is still localized in the excited miniband at 
$t=T_B = 2\pi/F$.
In this parameter range, the dynamics is very sensitive even to
small nonlinearities as shown in Fig.~\ref{fig-bzo1} (d).
The survival probability at $t=T_B$ differs 
significantly for $UN = -1$ and $UN=+1$, although the nonlinearity 
is still comparatively weak.

For the given interaction strength $|U| \le 1$, the splitting and also the 
recombination of the condensate is fully coherent; the fraction of 
non-condensed atoms is less than $0.4 \%$ at $t=2T_B$ as shown 
in  Fig.~\ref{fig-bzo1} (a).
A significant depletion of the condensate is observed only for stronger
nonlinearities; for instance the non-condensed fraction at $t=2T_B$ 
exceeds $10 \%$  for $UN \apprge 5$. 

This example demonstrates the possible use of Landau-Zener 
tunneling and Bloch-Zener oscillations in quantum metrology.  
These tools can be used, for instance, to directly measure the band 
structure of a bichromatic potential as demonstrated in \cite{Klin10}. 
This is a unique feature of bichromatic optical lattices. In a simple 
periodic potential, a matter wave will be accelerated further towards 
$-\infty$ after it has escaped from the ground band, such that no 
interference can be observed.

\subsection{Coupling of bands by a periodic driving}

Previously, we have discussed the effects of Zener tunneling between the
two minibands induced by the external field $F$. A coupling of the bands
can also be introduced in the field free case by a periodic driving 
of the system parameters. This has the advantage that the quasi 
momentum $\kappa$ is conserved such that a different regime of 
the dynamics can be explored.

\begin{figure}[t]
\centering
\includegraphics[width=8.5cm,  angle=0]{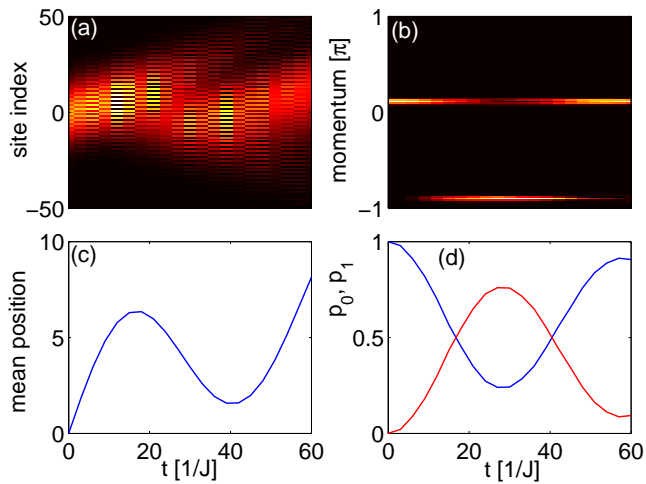}
\caption{\label{fig-dirac1}  
(Color online)
Quantum dynamics of BEC in a bichromatic lattice with periodic driving 
$\delta(t)$ and $F=0$. 
(a,b) Atomic density in real and momentum space, respectively.
(c) Position expectation value $\sum \nolimits_j j \langle \hat n_j \rangle/N$. 
(b) Occupation of the two minibands $p_{0}$ (blue) and $p_1$ (red).
}
\end{figure}

Here we consider a BEC initially prepared in the ground miniband with a
well defined quasimomentum $\kappa$. The strength of the 
double-periodic optical lattice is varied in time to realize a harmonic
driving of the energy offset
\be
  \delta(t) = \delta_0 + \delta_1 \cos(\omega t).
\ee
This driving induces transitions between the two minibands if the 
frequency is chosen to be  resonant with the band gap, 
$\omega = E_1(\kappa,\delta_0) - E_0(\kappa,\delta_0)$. In the following 
example we set $\delta_0 = 0.4$, $\delta_1 = 0.2$ and $U = 0$. The initial 
state is assumed to be pure BEC with momentum $\kappa = 0.1\pi$, 
weighted by a Gaussian envelope with $\sigma = 10$. 
The resulting dynamics is 
shown in Fig.~\ref{fig-dirac1} in real (left) and momentum space (right).
One clearly observes the transitions between the two minibands,
while the quasimomentum of the BEC is conserved (panel (b)). 
A further quantitative analysis of this effect is provided in panel (d), 
where the occupation of the two minibands $p_{0,1}$ is plotted.
The oscillation between the bands has remarkable consequences for 
the real-space dynamics of the BEC shown on the left-hand side of the 
figure. As the two minibands have opposite curvature, a transition 
between the bands reverses the group velocity of the matter wave. This 
leads to an oscillatory motion of the mean position, which can be 
understood as a quantum simulation of the {\em Zitterbewegung} 
of a Dirac spinor. This relativistic effect results from the interference 
of particle and anti-particle contributions moving to opposite directions. 
In the discussed quantum simulator, the two minibands thus play the 
role of particle and anti-particle contributions, respectively.
Similar effects were recently predicted for optical waveguide 
arrays \cite{Long10a,Long10b}.

\section{Conclusion and Outlook}

We have discussed the dynamics of a Bose-Einstein condensate
in a bichromatic optical lattice. In such a lattice, the ground  
Bloch band splits up into two minibands with a controllable
band gap. Hence, they are ideally suited to study the complex
quantum dynamics resulting from the interplay of the intraband 
dynamics and transition between the minibands. 

The basic dynamics of a BEC in an optical lattice can be inferred
from the band structure of the system. Within the tight-binding
approximation, one can readily calculate the linear as well
as the nonlinear Bloch states. In particular, this treatment 
yields an explicit expression for the critical interaction strength 
for the occurrence of looped Bloch bands, which leads to a 
breakdown of adiabatic motion. The stability properties of 
the Bloch states have been analyzed in detail by the
Bogoliubov-de Gennes approach.

The dynamics of a BEC was simulated using the 
Bogoliubov backreaction method which also provides a quantitative
estimate for the depletion of the condensate.
The Landau-Zener tunneling of a BEC between minibands in 
a tilted or accelerated bichromatic lattice has been intestigated in 
detail. For weak interactions, the condensate remains essentially
intact, whereas the Landau-Zener tunneling process is strongly
affected. Repulsive interactions increase the tunneling rate in particular
in the 'adiabatic regime' of large band gaps, while attractive
interactions suppress Zener tunneling. Strong interactions 
cause a dynamical instability and thus a rapid depletion of 
the condensate mode. However, the mechanism of dynamical 
instability is significantly different: In the repulsive case, instability sets in at 
the edge of the Brillouin zone and is intimately related to the occurrence of 
looped Bloch bands. A condensate with attractive interactions is unstable
already in the center of the Brillouin zone, leading to a collapse
and explosion of the condensate.

On longer timescales, the interplay of Bloch oscillations and Landau-Zener
tunneling leads to a complex dynamics due to the interference of the contributions
in the two minibands. A quantitative analysis of this effecs has been
given, which also shows the applicability of Bloch-Zener oscillations
in matter wave interferometry.
A remarkable dynamics is also observed if transitions between the bands are 
not induced by a static external field, but by a periodic driving which leaves the
quasimomentum unchanged. Because of the different curvature of the 
minibands, the matter waves in the two minibands move into opposite directions. 
The interference of the two contributions then leads to a dynamics which is 
comparable to the {\em Zitterbewegung} of a Dirac spinor.

\begin{acknowledgments}

This work has been supported by the German Research Foundation (DFG) 
through the research fellowship program (grant no WI 3415/1-1) and the 
Graduiertenkolleg 792 as well as the Studienstiftung des deutschen Volkes.
We thank M.~Weitz, T.~Salger and M.~Wubs for inspiring discussions.
\end{acknowledgments}


\begin{thebibliography}{10}

\bibitem{Bloc28}
F.~Bloch,
Z. Phys. {\bf 52}, 555 (1928). 

\bibitem{Daha98}
M.~Ben Dahan, E.~Peik, J.~Reichel, Y.~Castin, and C.Salomon,  
Phys. Rev. Lett. {\bf 76},  4508 (1996).

\bibitem{Land32b}
L.~D. Landau,
Phys.~Z.~Sowjetunion {\bf 2}, 46 (1932).

\bibitem{Zene32}
C.~Zener,
Proc.~R.~Soc.~London {\bf 137}, 696 (1932).

\bibitem{Majo32}
E.~Majorana,
Nuovo~Cimento {\bf 9}, 43 (1932).

\bibitem{Stue32}
E.~C.~G. St{\"u}ckelberg, 
Helv.~Phys.~Acta {\bf 5}, 369 (1932).

\bibitem{Legg01}
A.~J. Leggett,
Rev.~Mod.~Phys. {\bf 73}, 307 (2001).

 

\bibitem{Mors06}
O.~Morsch and M.~K. Oberthaler,
Rev. Mod. Phys. {\bf 78}, 179 (2006).


\bibitem{Gorl01}
A.~G\"orlitz, T.~Kinoshita, T.~W. H\"ansch, and A.~Hemmerich,
Phys. Rev. A {\bf 64}, 011401(R)  (2001).

\bibitem{Foll07}
S.~F\"olling, S.~Trotzky, P.~Cheinet, M.~Feld, R.~Saers, 
A.~Widera, T.~Mueller, and I.~Bloch,
Nature {\bf 448}, 1029 (2007).

\bibitem{Ritt06}
G.~Ritt, C.~Geckeler, T.~Salger, G.~Cennini, and M.~Weitz, 
Phys.~Rev.~A {\bf 74}, 063622 (2006),

\bibitem{Salg07}
T.~Salger, C.~Geckeler, S.~Kling, and M.~Weitz,
Phys. Rev. Lett. {\bf 99}, 190405 (2007). 

\bibitem{Salg08}
T.~Salger, G.~Ritt, C.~Geckeler, S.~Kling, and M.~Weitz, 
Phys. Rev. A {\bf 79}, 011605(R) (2009).

\bibitem{Salg09}
T.~Salger, S.~Kling, T.~Hecking, C.~Geckeler, L.~Morales-Molina, and M.~Weitz,
Science {\bf 326}, 1241 (2009).

\bibitem{Klin10}
S.~Kling, T.~Salger, C.~Grossert, and M.~Weitz,
Phys. Rev. Lett. {\bf 105}, 215301 (2010)

\bibitem{Wu00}
B.~Wu and Q.~Niu, 
Phys.~Rev.~A {\bf 61}, 023402 (2000).

\bibitem{Zoba00}
O.~Zobay and B.~M. Garraway,
Phys.~Rev.~A {\bf 61}, 033603 (2000).

\bibitem{Liu02}
Jie Liu, Libin Fu, Bi-Yiao Ou, Shi-Gang Chen,
Dae-Il Choi, Biao Wu, and Qian Niu, 
Phys.~Rev.~A {\bf 66} 023404 (2002).

\bibitem{05level3}
E.~M. Graefe, H.~J. Korsch, and D.~Witthaut, 
Phys.~Rev.~A {\bf 73}, 013617 (2006).

\bibitem{Ande98}
B.~P. Anderson and M.~A. Kasevich, 
Science {\bf 282}, 1686 (1998).

\bibitem{Lasi03}
M.~Jona-Lasinio,  O.~Morsch, M.~Cristiani, N.~Malossi, 
J.~H. M\"uller, E.~Courtade, M.~Anderlini, and E.~Arimondo, 
Phys.~Rev.~Lett. {\bf 91}, 230406 (2003).

\bibitem{Fall04}
L.~Fallani, L.~De Sarlo, J.~E. Lye, M.~Modugno, 
R.~Saers, C.~Fort, and M.~Inguscio, 
Phys.~Rev.~Lett. {\bf 93}, 140406 (2004).

\bibitem{Sias07}
C.~Sias, A.~Zenesini, H.~Lignier, S.~Wimberger, D.~Ciampini, 
O.~Morsch, and E.~Arimondo,
Phys. Rev. Lett. {\bf 98}, 120403 (2007).


\bibitem{Zene09}
A.~Zenesini,  H.~Lignier, G.~Tayebirad, J.~Radogostowicz, 
D.~Ciampini, R.~Mannella, S.~Wimberger, O.~Morsch, and E.~Arimondo, 
Phys. Rev. Lett. {\bf 103}, 090403 (2009). 


\bibitem{06zener_bec}
D.~Witthaut, E.~M. Graefe, and H.~J. Korsch ,
Phys. Rev. A {\bf 73}, 063609 (2006).

\bibitem{Wu06}
Biao Wu and Jie Liu,
Phys. Rev. Lett. {\bf 96}, 020405 (2006).

\bibitem{10zener_phase}
F.~Trimborn, D.~Witthaut, V.~Kegel, and H.~J. Korsch,
New J.~Phys. {\bf 12}, 053010 (2010).


\bibitem{06bloch_zener}
B.~M. Breid, D.~Witthaut, H.~J. Korsch,  
New~J.~Phys. {\bf 8}, 110 (2006).

\bibitem{06bloch_manip}
B.~M. Breid, D.~Witthaut, H.~J. Korsch,  
New~J.~Phys. {\bf 9}, 62 (2007).


\bibitem{Jaks98}
D.~Jaksch, C.~Bruder, J.~I. Cirac, C.~W. Gardiner, and P.~Zoller,
Phys. Rev. Lett. {\bf 81}, 3108  (1998).

\bibitem{Vard01}
A.~Vardi and J.~R. Anglin,
Phys. Rev. Lett. {\bf 86}, 568 (2001).

\bibitem{Angl01}
J.~R. Anglin and A.~Vardi, 
Phys. Rev. A {\bf 64}, 013605 (2001). 

\bibitem{Tikh07}
I.~Tikhonenkov, J.~R. Anglin and A.~Vardi, 
Phys. Rev. A {\bf 75}, 013613 (2007).


\bibitem{09dcomb}
D.~Witthaut, K.~Rapedius, and H.~J. Korsch, 
J. Nonlin. Math. Phys. {\bf 16}, 207 (2009).

\bibitem{Wu03}
B.~Wu and Q.~Niu,
New~J.~Phys. {\bf 5}, 104 (2003).

\bibitem{Cast98}
Y.~Castin and R.~Dum,
Phys. Rev. A {\bf 57}, 3008 (1998).


\bibitem{07phase}
F.~Trimborn, D.~Witthaut, and H.~J.~Korsch, 
Phys.~Rev.~A {\bf 77}, 043631 (2008).


\bibitem{09phase_appl}
F.~Trimborn, D.~Witthaut, and H.~J.~Korsch, 
Phys.~Rev.~A {\bf 79}, 013608 (2009).

\bibitem{Kolo09}
A.~R. Kolovsky, H.~J. Korsch, and E.-M. Graefe, 
Phys. Rev. A {\bf 80}, 023617 (2009). 

\bibitem{05bloch_bec}
D.~Witthaut, M.~Werder, S.~Mossmann, and H.~J.~Korsch,
Phys.~Rev.~E {\bf 71}, 036625 (2005).



\bibitem{Gust08}
M.~Gustavsson, E.~Haller, M.~J. Mark, J.~G. Danzl, 
G.~Rojas-Kopeinig, and H.-C. N\"agerl,  
Phys. Rev. Lett. {\bf 100}, 080404 (2008). 

\bibitem{Long10a}
S.~Longhi, Phys. Rev. B {\bf 81}, 075102 (2010).

\bibitem{Long10b}
S.~Longhi, Phys. Rev. A {\bf 81}, 022118 (2010).





\end{thebibliography}
\end{document}